\documentclass{article}
\usepackage[dvips]{graphicx}
\usepackage{natbib}
\usepackage{amssymb}
\usepackage{a4wide}

\newcommand\nodata{ ~$\cdots$~ }
\newcommand\meai{\buildrel\hbox{$\scriptstyle <$}\over{\lower 1pt
\hbox{$\scriptstyle\sim$}}}

\date{May 22, 2001}

\title{Cometary activity in 2060 Chiron\\
 at minimum brightness\\[15pt]
{\small (Accepted for publication in Planetary \& Space Science)}}

\author{Adri\'an M. Silva
\footnote{Instituto de Astronom\'{\i}a y F\'{\i}sica del Espacio,
 C. C. 67 Suc.\ 28, 1428, Buenos Aires, Argentina}
\footnote{Visiting Astronomer, Complejo Astron\'omico El Leoncito,
operated under agreement between the Consejo Nacional de
Investigaciones Cient\'{\i}ficas y T\'ecnicas de la Rep\'ublica
Argentina and the Universities of La Plata, C\'ordoba, and San Juan.}
\and
Sergio A. Cellone
\footnote{Facultad de Ciencias Astron\'omicas y Geof\'{\i}sicas, Universidad
 Nacional de La Plata, Paseo del Bosque, 1900 La Plata, Argentina}~$^\dag$
}

\begin{document}
\maketitle

\begin{abstract}
We present two-colour CCD imaging of 2060 Chiron obtained between 1996
and 1998 with the 2.15\,m telescope at CASLEO (San Juan,
Argentina). These post-perihelion observations show that Chiron was
then near its historical brightness minima, however a coma was clearly
detected. The dynamical state of the coma is studied by means of
azimuthally averaged surface brightness profiles, which show the
signatures of radiation pressure on the dust grain distribution.
Aperture photometry shows an achromatic dimming with an amplitude
$\approx 0.09$ mag in approximately one hour. If due to rotation of
the nucleus, this rather high amplitude is used to derive a new value
for the nuclear magnitude, $m_0 \approx 6.80$ mag.

\end{abstract}

\section{Introduction}

The Centaur (2060) Chiron has been the target of extensive study,
particularly since 1989 when its cometary activity was first detected
\citep{MeBe89}. Ground-based
surface photometry revealed a dust coma extending out to $\approx 2
\times 10^5$ km from the nucleus \citep{LJ90,MeBe90,W91,MaBu93,Luu93},
while the inner coma was probed by means of stellar occultations
\citep{Elliot95,Bus96} and using HST \citep{Meech97}. Water ice has
been recently detected in Chiron \citep{Foster99,LJT00}, and its
cometary activity is supposed to be powered by the release of trapped
$CO$ within the $H_2O$ ice matrix \citep{PBI95,FS97}.

Photometric observations have revealed, in addition to the rotational
light\-curve, brightness variations with time-scales of days, weeks,
and years which are ascribed to changing activity. However, it
remains to be explained why Chiron's brightness decreased while
approaching perihelion, and why the presence of a prominent coma is
not always coincident with brightness maxima \citep[e.g.,][and
references therein]{Lazz97}. Hence, a continuous monitoring of this
object, especially near perihelion, has often been
encouraged. Considered as a nearby representative of Kuiper Belt
objects, investigation of Chiron, and Centaurs in general, is ideal
for getting a closer look at the frozen bodies of the outer Solar
System \citep[e.g.,][]{SC96}. In this paper we present broad-band CCD imaging
of Chiron during a post-perihelion period (1996--1998) when published
optical photometry is rather scarce. The observations are described in
Section \ref{s_obs}, while Section \ref{s_rd} presents the results of
our aperture and surface photometry. Our conclusions are given in
Section \ref{s_conc}.

\section{Observations}
\label{s_obs}
We obtained broad--band $V$ (Johnson) and $R$ (Cousins) CCD imaging of
2060 Chiron on 5 nights during three different observing runs with the
``Jorge Sahade'' 2.15\,m telescope of CASLEO (San Juan, Argentina).
The detector was a cryogenically cooled Tek 1024 CCD, with a gain of
1.98 electrons adu$^{-1}$ and a read-out noise of 9.8 electrons. A focal
reducer provided a field of view approximately 9 arcmin in diameter,
with a scale of 0.82 arcsec pix$^{-2}$.
 The telescope was tracked at sidereal rate, hence exposure times of
individual frames were limited to a maximum of 180 sec in order to
avoid smearing of the images because of Chiron's proper motion
($\lesssim 0.2$ arcsec min$^{-1}$).

The observations span a time interval of 16 months; the dates, UT at
mid-exposure, and individual exposure times at $V$ and $R$,
respectively, are given in the first four columns of
Table~\ref{t_aph}. Weather conditions varied along the nights, with
very good atmosphere transparency during the 1997 run, but with thin
cirrus during the two other runs, while seeing FWHM was never better
than $\sim 2.3$ arcsec. Several standard stars fields from \citet{L92} were
also observed each night for calibrating the photometry to the
standard system.  Data reduction was made following normal procedures
with the {\sc iraf} software package\footnote{IRAF is distributed by
the National Optical Astronomy Observatories, which are operated by
the Association of Universities for Research in Astronomy, Inc., under
cooperative agreement with the National Science Foundation.}. All
frames were debiased and flat-fielded using high signal--to--noise
dome flats.


\section{Results and Discussion}
\label{s_rd}

\subsection{Aperture Photometry}
\label{s_apphot}

Aperture photometry of Chiron and the standard stars was performed on
every individual image with the {\sc apphot} package within {\sc
iraf}, using an aperture radius of 10 pix (8.2 arcsec). Given the
relatively high galactic latitude of Chiron during the observations
($38^\circ \le b \le 52^\circ$), there were no severe crowding
problems in our images.

Table \ref{t_aph} also gives the apparent $V$ magnitude and $V-R$ colour,
along with their respective photometric errors, for each
observation. The corresponding heliocentric ($r$) and geocentric
($\Delta$) distances and phase angles ($\alpha$) were taken from the
ephemeris, and were used to calculate the absolute ($H_V$) magnitudes,
following \citet{Bowell_89}, and adopting a slope parameter $G=0.70$
\citep{Bus_89} to correct the magnitudes to phase angle $\alpha =
0^\circ$. However, the presence of some degree of cometary activity
(see Sect.~\ref{s_surfphot}) probably invalidates the use of such $G$
value \citep{Lazz97}, hence reduced ($H_{V(\alpha)}$, i.e., without
any phase correction) magnitudes are also listed. Apparent magnitudes
and astrometry obtained from the same observations have been reported
by \citet{OCS98}.

\begin{figure}[h]
\begin{center}
\includegraphics[width=0.62\textwidth]{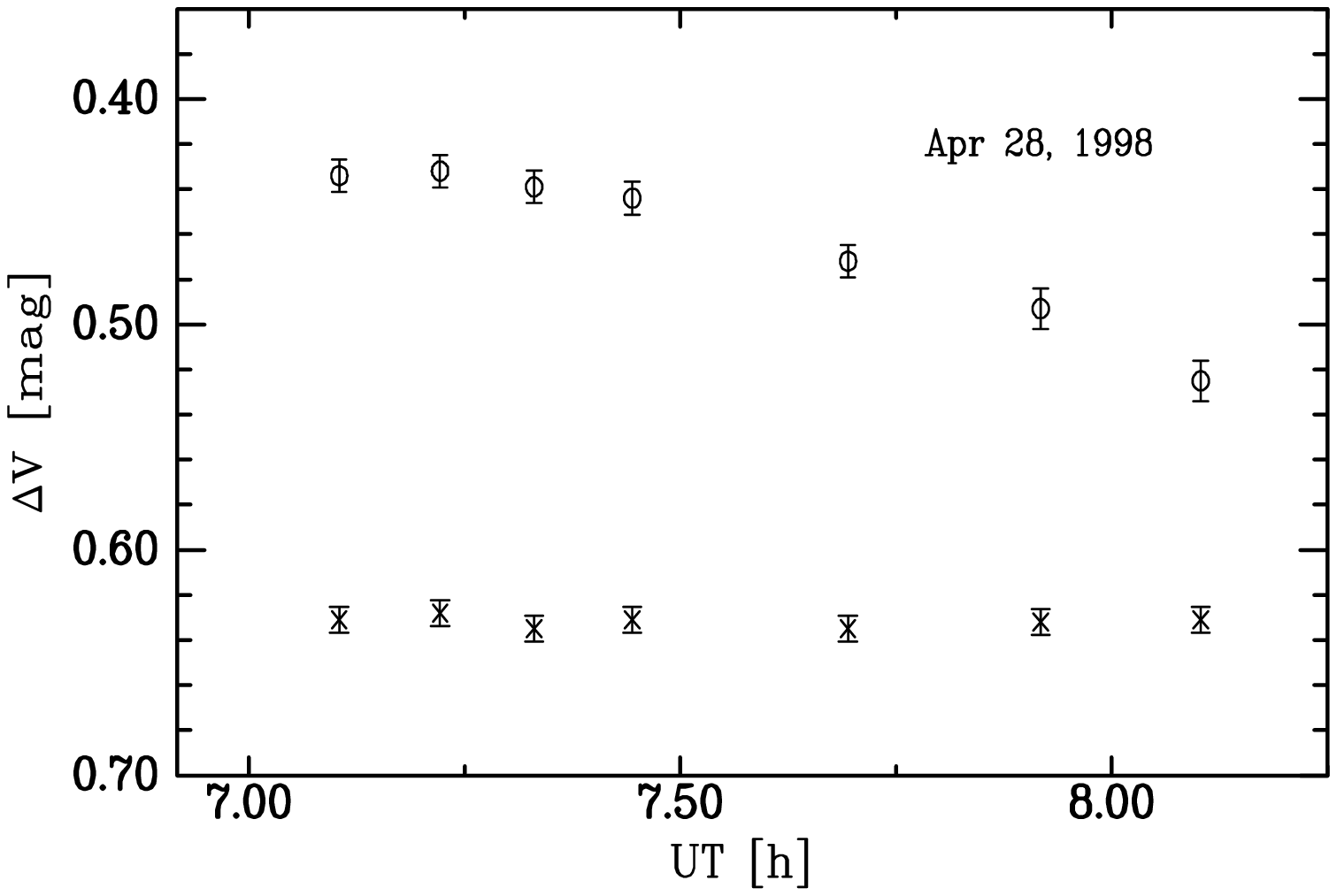}
\end{center}
\caption{Differential magnitude Chiron $-$ comparison field star
$S_1$ (open circles) and star $S_2$ $-$ star $S_1$ (crosses) as a
function of UT, for Apr 28, 1998. An arbitrary vertical
offset was applied to the $S_2-S_1$ data.}
\label{f_mdif98}
\end{figure}

During our 1997 and 1998 observations Chiron was very close to its
historical minima ($H_V \approx 6.8$ mag) of 1983-85 and 1995
\citep[and references therein]{Lazz96}, while in late 1996 it appeared
slightly brighter, although the rather large error bar in this case
prevents any reliable comparison.
 Mean absolute $V$ magnitudes were $\langle H_V \rangle= 6.74$ and
$\langle H_V \rangle= 6.67$ for the 1997 and 1998 runs, respectively.
The 1998 observations were done under non-photometric conditions,
hence they had to be tied to a previous night with better weather
through 6 stars within the CCD field that were observed both
nights. Although a systematic error may thus still be present in the
absolute photometry, a differential $V$ lightcurve for Chiron against
a comparison field star ($S_1$) clearly shows a $0.09$ mag dimming in
about one hour, as displayed in Figure~\ref{f_mdif98}. Differential
magnitudes ($\Delta m$) of a second control field star ($S_2$) against
$S_1$, also shown, give a stability check.  After correcting for
light-time effect and adopting a synodic period of 5.917813~hs
\citep{MaBu93}, this brightness decrease is consistent with Chiron's
rotational lightcurve \citep{LJ90, Lazz97} although the peak-to-peak
amplitude seems to be larger than most published values ($\Delta m
\lesssim 0.06$ mag).  Our result matches the amplitude observed by
\citet{Bus_89} during the 1986 brightness minimum, when no coma was
directly observed and, hence, a small dilution was expected.  Note
that, if the brightness variation we report here is purely rotational,
we can only set a lower limit to its amplitude since a clear minimum
was not detected.

The corresponding $V-R$ colours in Table~\ref{t_aph} get slightly
bluer with time; however, this bluing is only apparent, being caused
by the fact that the $R$ frames were taken a few minutes after the
corresponding $V$ frames, and hence Chiron was then dimmer. Correction
of this effect (by interpolating the $R$ magnitudes at the same
instants of observation as the $V$ ones) gives a fairly constant
colour, with a mean $\langle V-R \rangle = 0.446 \pm 0.004$. This
achromatic behaviour supports rotation of an asymmetric object as the
origin of the light variation, against the effect of dark (and red)
spots on the body's surface.

Following \citet[see their Eq.\ 19]{LJ90}, a bare nucleus magnitude
$m_0$ can be estimated from the relation between observed light-curve
amplitude ($\Delta m_i$) and mean absolute magnitude. Our data
($\Delta m_i \ge 0.09$ mag, $\langle H_V \rangle = 6.67$) are best
fitted with a relatively faint nuclear magnitude $m_0 \approx 6.80$
\citep[see also][]{MaBu93} and a bare nucleus amplitude $\Delta m_0
\approx 0.10$ mag. However, Lazzaro et al.'s (1997) data are
consistent with a much lower $\Delta m_0$. An alternative explanation
could be that the 1998 dimming was part of a larger, non-rotational
brightness variation.

The 1997 data, obtained under photometric conditions, also show a
$\sim 0.06$ mag dimming from the first to second nights, although in
this case the time sampling is insufficient for trying a match with
the rotational light-curve.


\subsection{Surface Photometry}
\label{s_surfphot}

Azimuthally averaged surface brightness profiles (SBPs) of Chiron and
two comparison stars within the same CCD frame were obtained with the
{\sc ellipse} task within {\sc iraf - stsdas} from our deepest images,
those obtained on 1997, June 27, and 1998, April 28. Several (2--7)
frames per band corresponding to each night were summed up in order to
improve the signal-to-noise ratio. Background objects were masked-out
and a clipping routine was used to eliminate remaining faint objects
and bad pixels. Comparison stars with magnitudes as close as possible
to that of Chiron, and lying within no more than $\sim 2$ arcmin of
projected distance from it, were always selected.

The images obtained on Apr 28, 1998 span a considerable range in time
($\sim 1$ hour), hence two different composite frames were generated
for each band: one using the centroids of several field stars for
registering the frames before summing them up, and the other using the
centroids of Chiron's images. This procedure thus resulted in two sets
of images, one with sidereal tracking, from which the comparison stars
SBPs were obtained, and the other set as if ``coarsely tracked'' on
Chiron, with 840 and 700 sec total exposure times in $V$ and $R$,
respectively.

\begin{figure}[t]
\begin{center}
\hbox{
\includegraphics[width=0.5\textwidth]{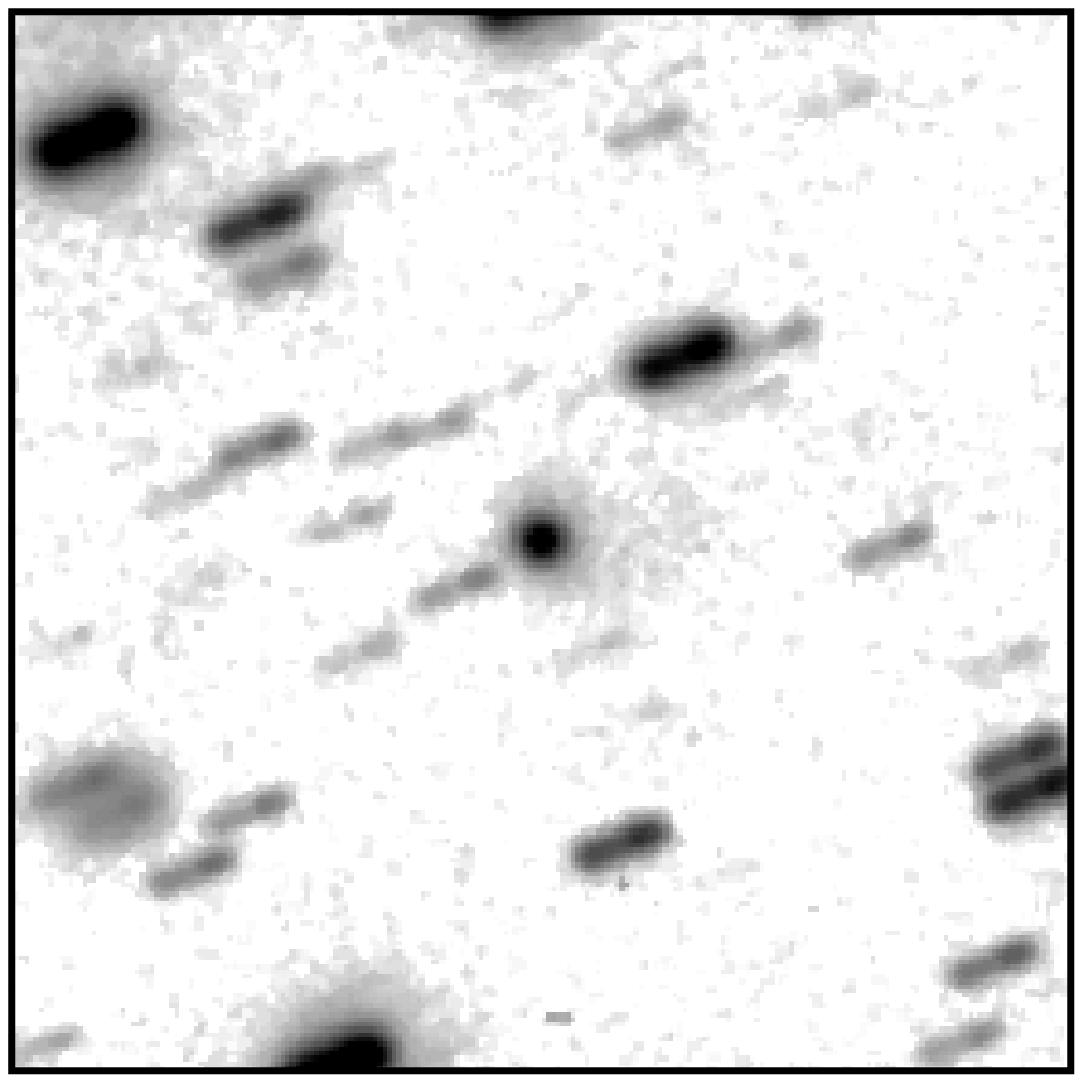}\hskip 25pt
\includegraphics[width=0.5\textwidth]{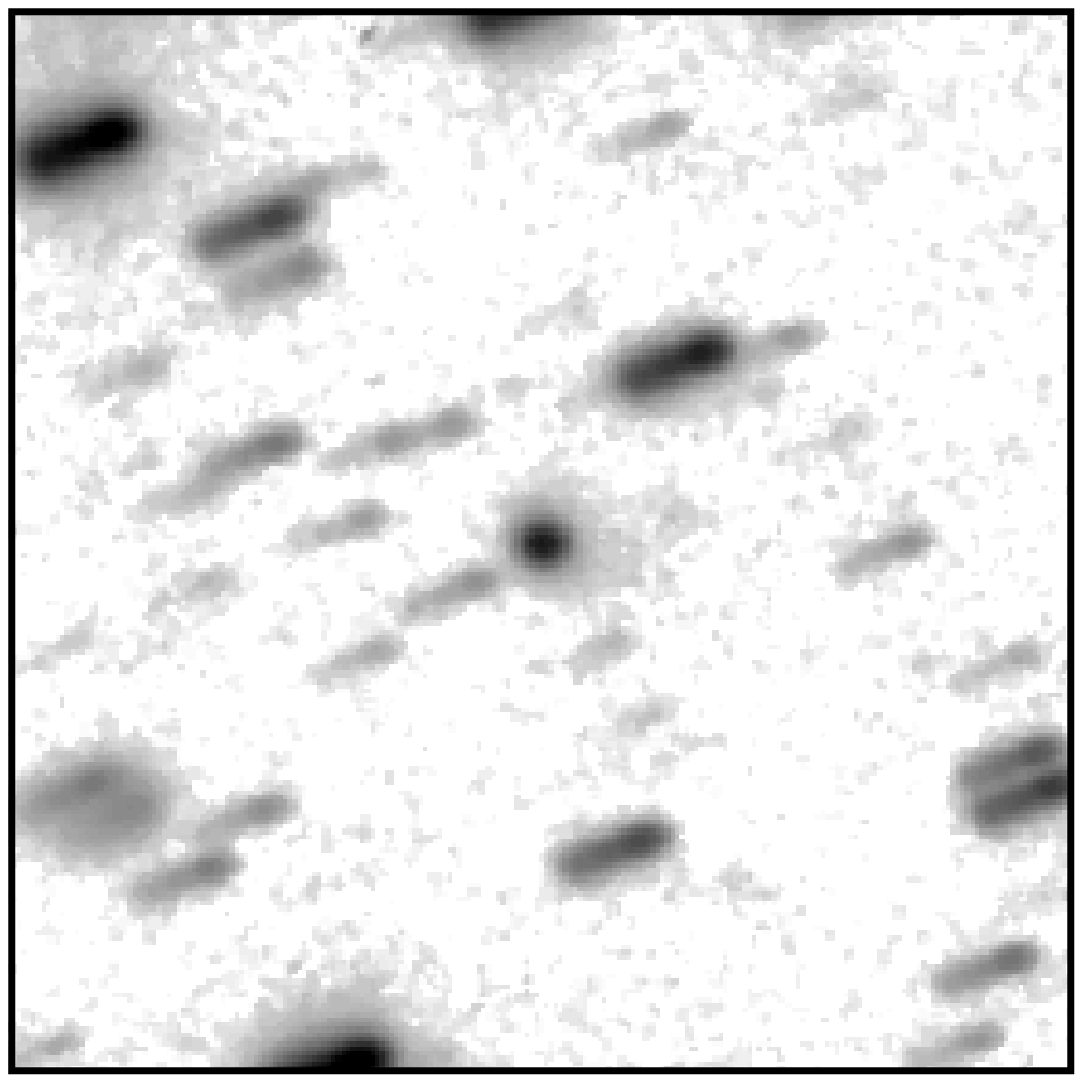}
}
\end{center}
\caption{Grey scale reproductions of $V$ (left) and $R$ (right)
composite images ``tracked'' on 2060 Chiron (Apr 28, 1998), with sky
subtracted. A median filter was applied to improve presentation, and
intensities were transformed to mag arcsec$^{-2}$. Each frame is
$3\times 3$ arcmin, with North up and East to the left. The anti-solar
direction is at position angle $265^\circ$.}
\label{f_grisVyR}
\end{figure}

\begin{figure}[hp]
\begin{center}
\hbox{
\includegraphics[width=0.5\textwidth]{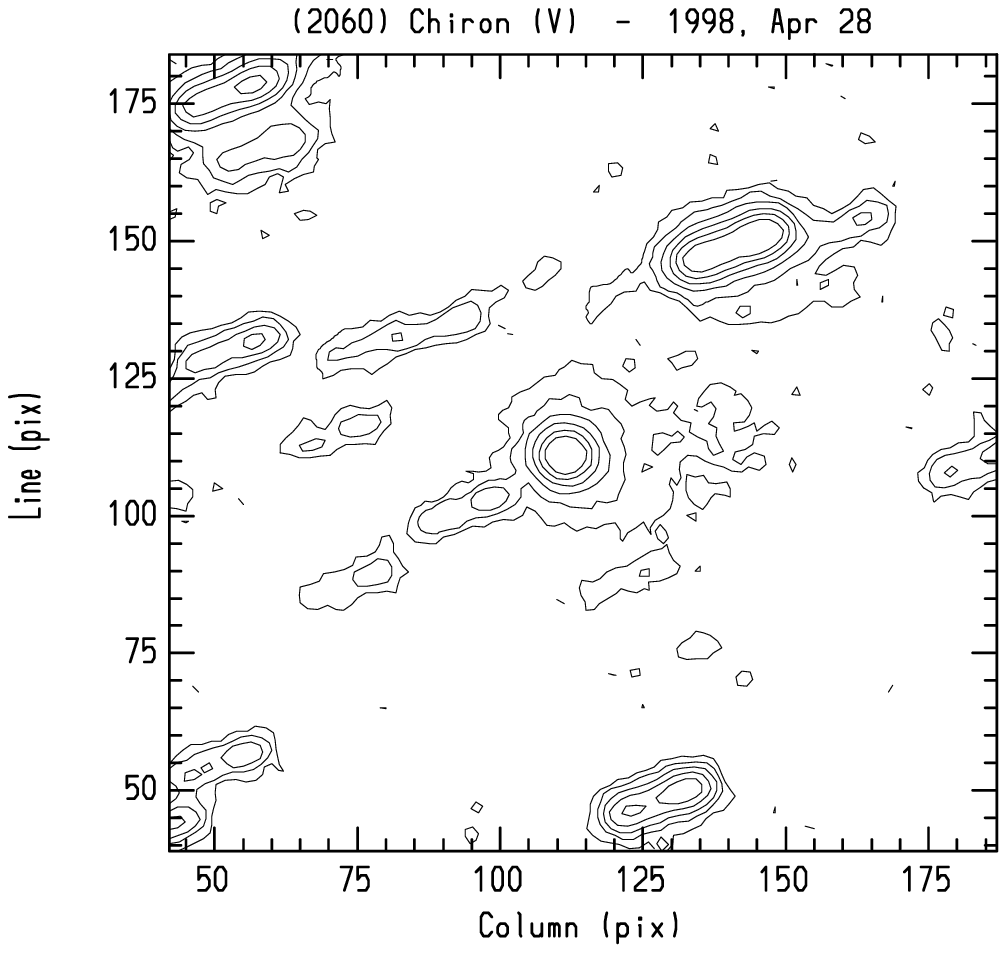}\hskip 25pt
\includegraphics[width=0.5\textwidth]{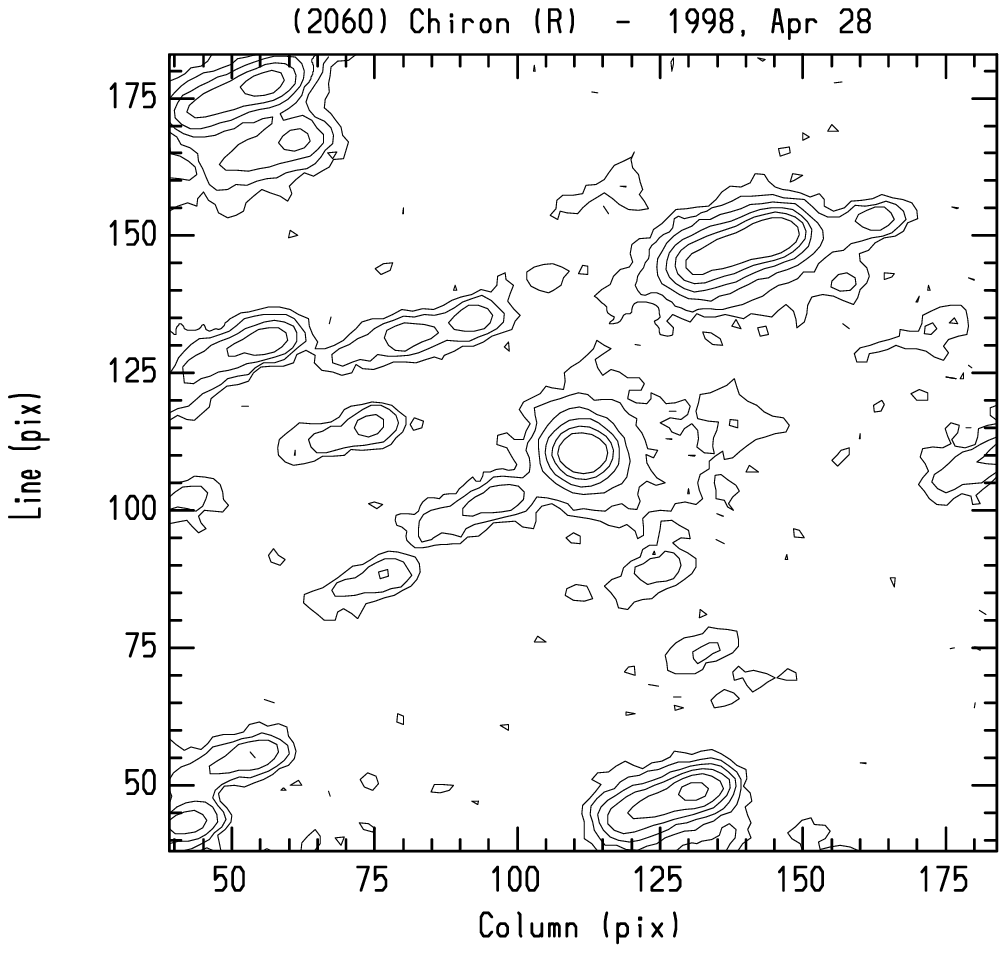}
}
\end{center}
\caption{Contour plots of the same images of Fig.~\ref{f_grisVyR}, but
showing the central $2\times 2$ arcmin field. The faintest contours
displayed correspond to $\mu_V=26.2$ mag arcsec$^{-2}$ (left) and
$\mu_R=26.0$ mag arcsec$^{-2}$ (right), respectively; adjacent
contours are spaced by 1 mag arcsec$^{-2}$.}
\label{f_contVyR}
\end{figure}

Grey level reproductions of the central $3\times 3$ arcmin of the
Chiron--tracked, sky-subtracted images are shown in
Figure~\ref{f_grisVyR}. North is up and East to the left, with the
anti-solar direction at position angle $265^\circ$. A faint coma
extending at least $\sim 18$ arcsec ($1.02\times 10^5$ km projected
distance) from the centre of Chiron's image is visible in that
direction. Note that Chiron's orbital motion was directed at position
angle $288^\circ$, as can be seen from the star trails. Contour plots
(of $2\times 2$ arcmin sub-images) are also shown in
Figure~\ref{f_contVyR}, with the faintest contours displayed
corresponding to $\mu_V=26.2$ mag arcsec$^{-2}$ and $\mu_R=26.0$ mag
arcsec$^{-2}$, respectively, and a 1 mag arcsec$^{-2}$ contour
spacing.

\begin{figure}[hp]
\begin{center}
\hbox{
\includegraphics[width=0.55\textwidth]{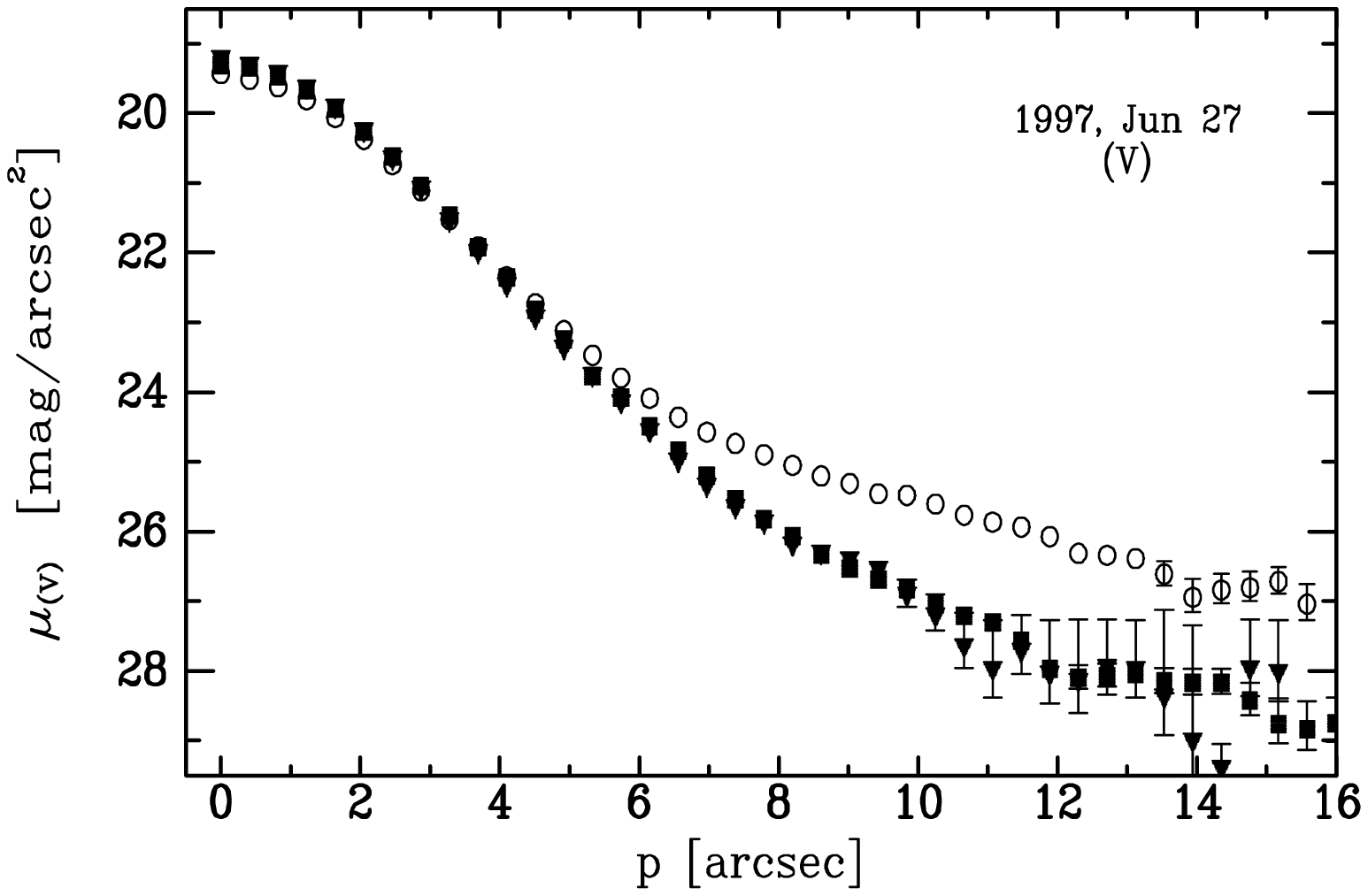}\hskip 10pt
\includegraphics[width=0.55\textwidth]{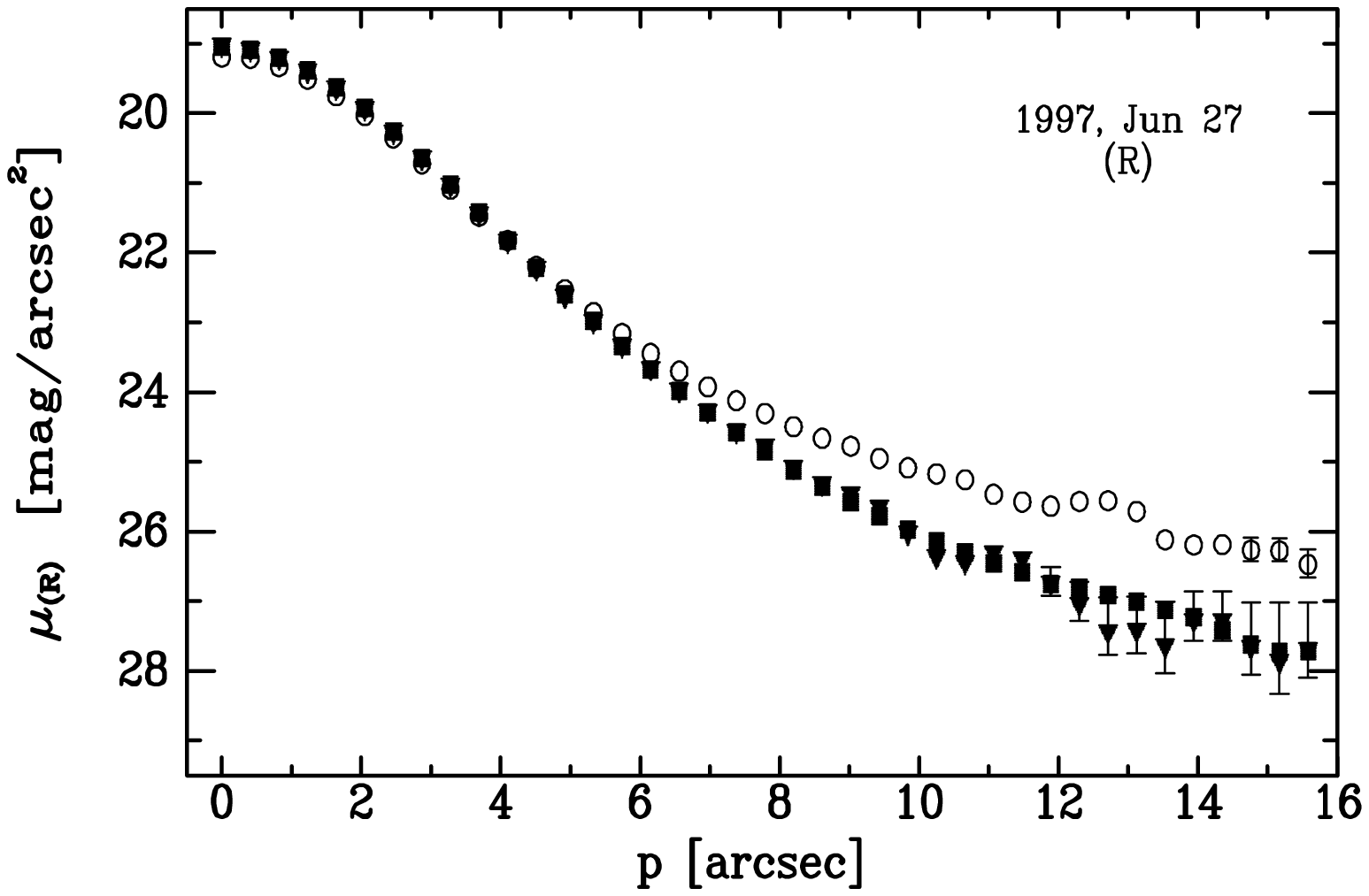}
}
\vskip 5pt
\hbox{
\includegraphics[width=0.55\textwidth]{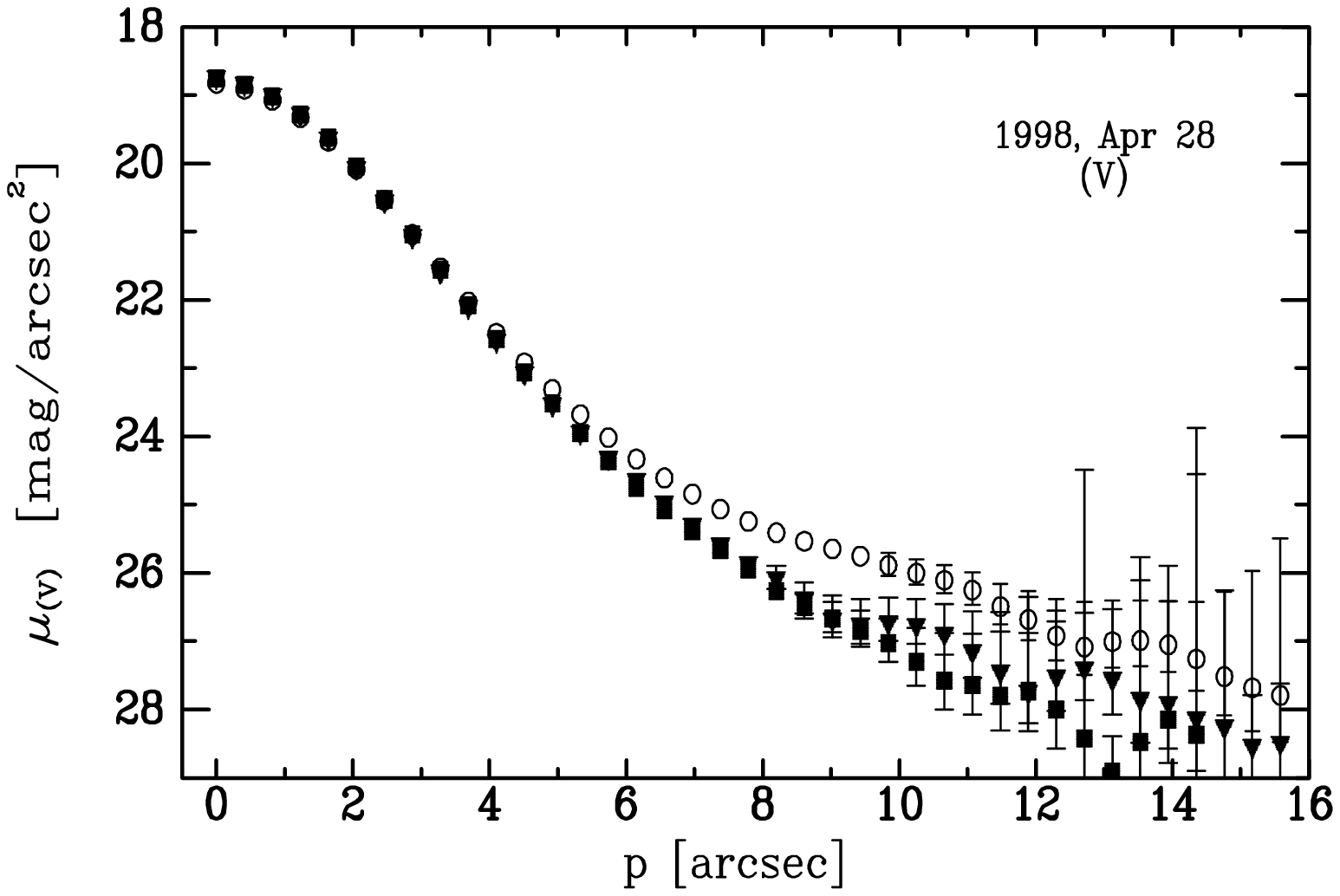}\hskip 10pt
\includegraphics[width=0.55\textwidth]{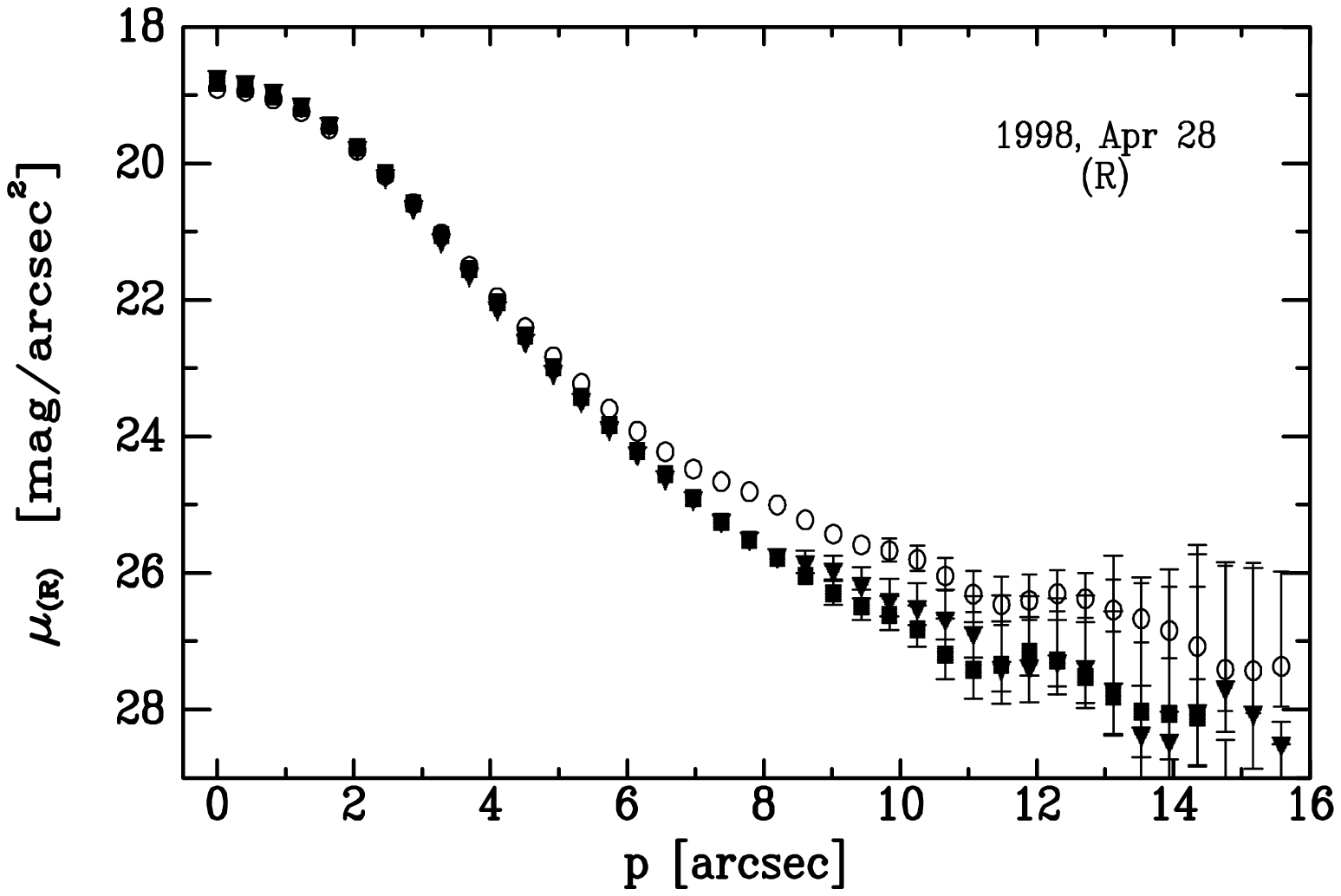}
}
\end{center}
\caption{Azimuthally averaged surface brightness profiles (SBPs) of
Chiron (open circles) and two field stars (filled symbols), scaled to
an identical integrated flux, from the co-added observations of Jun
27, 1997 (upper panels) and Apr 28, 1998 (lower panels).
 Error bars (not shown when smaller than the symbol size) represent
formal photometric errors plus those introduced by sky uncertainties.
Magnitudes are in the standard system, and each panel is labeled with
the corresponding filter.
}
\label{f_sbps}
\end{figure}

Chiron's coma is also evident from the high signal-to-noise SBPs as a
surface brightness enhancement above the stellar PSF for distances to
the nucleus $p \gtrsim 5''$ (Figure~\ref{f_sbps}); however, it is only
marginally detected in the images with shorter effective exposure
times obtained in 1996 and 1997 Jun 25-26. In every case, the fluxes
of the two comparison stars within an 8 arcsec radius aperture were
scaled to match that of Chiron. Special care was taken for sky
determination and subtraction, since low surface brightness features
are very sensitive to small errors in the adopted sky level. First, a
plane was fitted to the sky background of each image and this fit was
subtracted. Then, the SBPs were computed, and a final fine-tuning of
the local sky level for each object (typically a few adu) was made by
plotting the total flux vs.\ radius, and checking that the total flux
attained a constant value for sufficiently large radii ($\sim 1$
arcmin). Finally, the SBPs were re-computed after correcting for the
residual sky.

A power-law $S_{(p)} \propto p^s$, where $S_{(p)}$ is the surface
brightness in intensity units and $p$ is the radial distance from the
nucleus in arcsec, was fitted to the 1998 SBPs, after subtraction of
scaled mean stellar profiles. The fits were made in the log--log
plane, with the fitting range set to $6''\le p \le 11''$ ($1.79 \le
\ln p \le 2.40$), avoiding the seeing-smeared central regions as well
as the noisy outer regions. The slopes thus obtained were $s_R = -2.0
\pm 0.2$ and $s_V = -1.4 \pm 0.1$, for $R$ and $V$, respectively.

These values are roughly consistent with a dust coma were radiation
pressure effects are important \citep[$s = -1.5$; e.g.,][]{JM87}.
However, still allowing for the respective errors, the slope of the
SBP is steeper in $R$ than in $V$, corresponding to a change in colour
from $V-R=0.41 \pm 0.03$ at $p=5''$ to $V-R=0.05 \pm 0.16$ at
$p=11''$. Due to poorer seeing conditions, no reliable slope values
could be obtained from the 1997 data; however, a qualitatively similar
gradient is evident beyond the seeing disk and out to $p \simeq 12''$
($6.9 \times 10^4$ km) in a $\mu_{(V)} - \mu_{(R)}$ vs.\ $p$ plot.

This blue gradient could be tentatively explained as a scattering
effect due to a smaller mean grain size at larger distances from the
nucleus. Since the ratio of solar radiation pressure to solar gravity
on the grain ($\beta$) is inversely proportional to the grain radius
$a_\mathrm{g}$:
\begin{equation}
\beta = k {Q_\mathrm{rp} \over \rho_\mathrm{g} a_\mathrm{g}},
\end{equation}
where $k = 5.7398 \times 10^{-4}$ kg m$^{-2}$, $Q_\mathrm{rp}$ is the
radiation pressure scattering efficiency, and $\rho_\mathrm{g}$ is the
grain density \citep{Meech97}, smaller grains should be more
efficiently accelerated by radiation pressure thus traveling farther
from the nucleus than larger grains. 

 Curiously, this outwards bluing of Chiron's coma is opposite to the
gradient observed by \citet{W91} in 1990.  Since unnoticed systematic
errors may be affecting our photometry at very faint surface
brightness levels, independent confirmation would be desirable.

\section{Conclusions}
\label{s_conc}
Our post-perihelion observations ($8.65 \le r \le 8.94$ AU) show that
during 1997 and 1998 Chiron was still at a brightness minimum,
however, a coma was clearly detected. This coma can be traced out to a
projected distance $1.02\times 10^5$ km in the anti-solar direction,
and down to a surface brightness $\mu_V=26.2$ mag arcsec$^{-2}$ , i.e,
at least as faint as the deepest observations of Chiron published to
date \citep{Luu93}.

The slopes of the SBPs of this feature are consistent with a dust coma
affected by radiation pressure. On the other hand, a significant (at a
$\approx 3 \sigma$ level) gradient is detected, with $V-R$ colours
getting bluer outwards, both for 1997 and 1998 data. A grain
population with mean size decreasing with distance from the nucleus
could be invoked to explain the observed gradient, although, in the
light of opposite results \citep{W91}, our conclusion should be taken
with care and confronted with new observations.

Our aperture differential photometry, during the night with best time
coverage, shows that Chiron's brightness dimmed by $\approx 0.09$ mag
in roughly 1 hr without significantly changing its colour. If this
dimming is totally due to rotational effects (and not part of a
larger, non-asteroidal brightness change that our observations failed
to completely sample), the nuclear magnitude can be estimated at $m_0
\approx 6.80$ ($V$ band), with a bare nucleus light-curve amplitude
$\Delta m_0 \approx 0.10$ mag.

\section*{\normalsize \bf Acknowledgements}

Use of the CCD and data acquisition system supported under
U. S. National Science Foundation grant AST-90-15827 to R. M. Rich is
acknowledged. We thank Mario Melita for instructive discussion. We also
wish to thank Dr.\ K. Meech and a second (anonymous) referee for
useful comments.

\cleardoublepage
\section*{Figures}

\clearpage
\section*{Tables}

\begin{table}[h]
\begin{center}
\scriptsize
\caption{Aperture Photometry. \label{t_aph}}
\begin{tabular}{lcccccccccc}
\noalign{\vskip 5pt\hrule\vskip 1pt\hrule\vskip 3pt}
Date      &  UT    &$t_V$ &$t_R$&  $V$  & $V-R$  &  $r$   & $\Delta$ & $\alpha$ & $H_{V(\alpha)}$ & $H_V$ \\
          & (h:m) &(sec)&(sec)& (mag) & (mag)  &  (AU)  & (AU)  & (deg)  & (mag)   & (mag) \\
\noalign{\vskip 3pt\hrule\vskip 3pt}

1996\,Dec\,22 & 08:34 &\phantom{1}60&\nodata & $16.13 \pm 0.10$ & \nodata         & 8.530 & 8.956 & 5.813 & $6.71 \pm 0.10$ & $6.50 \pm 0.10$\\
1997\,Jun\,25 & 03:29 &\phantom{1}60&\nodata & $16.19 \pm 0.01$ & \nodata         & 8.645 & 8.212 & 6.257 & $6.93 \pm 0.01$ & $6.71 \pm 0.01$\\
            & 03:36 &\phantom{1}60&\phantom{1}40& $16.20 \pm 0.01$ & $0.45 \pm 0.04$ & 8.645 & 8.212 & 6.257 & $6.94 \pm 0.01$ & $6.72 \pm 0.01$\\
1997\,Jun\,26 & 02:13 &\phantom{1}20&\phantom{1}30& $16.25 \pm 0.01$ & $0.39 \pm 0.01$ & 8.645 & 8.226 & 6.296 & $6.99 \pm 0.01$ & $6.76 \pm 0.01$\\
            & 02:20 & \phantom{1}30&\phantom{1}30& $16.26 \pm 0.01$ & $0.39 \pm 0.01$ & 8.645 & 8.227 & 6.296 & $7.00 \pm 0.01$ & $6.77 \pm 0.01$\\
1997\,Jun\,27 & 01:45 &180&180& $16.24 \pm 0.01$ & $0.38 \pm 0.01$ & 8.646 & 8.242 & 6.334 & $6.98 \pm 0.01$ & $6.75 \pm 0.01$\\
            & 01:49 &180&180& $16.24 \pm 0.01$ & $0.40 \pm 0.01$ & 8.646 & 8.242 & 6.335 & $6.98 \pm 0.01$ & $6.75 \pm 0.01$\\
1998\,Apr\,28 & 07:06 &120&100& $15.96 \pm 0.02$ & $0.45 \pm 0.02$ & 8.938 & 7.941 & 0.974 & $6.70 \pm 0.02$ & $6.64 \pm 0.02$\\
            & 07:13 &120&100& $15.96 \pm 0.02$ & $0.43 \pm 0.02$ & 8.938 & 7.941 & 0.974 & $6.70 \pm 0.02$ & $6.64 \pm 0.02$\\
            & 07:20 &120&100& $15.96 \pm 0.02$ & $0.44 \pm 0.02$ & 8.938 & 7.941 & 0.973 & $6.70 \pm 0.02$ & $6.64 \pm 0.02$\\
            & 07:27 &120&100& $15.97 \pm 0.02$ & $0.43 \pm 0.02$ & 8.938 & 7.941 & 0.972 & $6.71 \pm 0.02$ & $6.65 \pm 0.02$\\
            & 07:42 &120&100& $16.00 \pm 0.02$ & $0.44 \pm 0.02$ & 8.938 & 7.941 & 0.971 & $6.73 \pm 0.02$ & $6.67 \pm 0.02$\\
            & 07:55 &120&100& $16.02 \pm 0.02$ & $0.43 \pm 0.02$ & 8.938 & 7.941 & 0.970 & $6.76 \pm 0.02$ & $6.70 \pm 0.02$\\
            & 08:06 &120&100& $16.05 \pm 0.02$ & $0.43 \pm 0.02$ & 8.938 & 7.941 & 0.969 & $6.79 \pm 0.02$ & $6.73 \pm 0.02$\\
\noalign{\vskip 3pt\hrule\vskip 1pt\hrule}
\end{tabular}
\end{center}
\end{table}


\end{document}